\documentclass{amsproc}
\newcommand{\sect}[1]{\section{#1}}
\newcommand{\subsect}[1]{\subsection{#1}}
\newcommand{\cen}{\]}
\newcommand{\ceq}{\[}

\newcommand{\nn}{\nonumber}

\newcommand{\blp}{\biggl(}
\newcommand{\brp}{\biggr)}
\newcommand{\lp}{\left(}
\newcommand{\rp}{\right)}
\newcommand{\lra}{\longrightarrow}
\newcommand{\prt}{\partial} 
\newcommand{\prtb}{\bar{\partial}}
\newcommand{\Cl}{\mbox{\Large $\displaystyle S$}}
\newcommand{\Llra}{\ \Longleftrightarrow\ }
\newcommand{\Lra}{\ \Longrightarrow\ }
\newcommand{\lbl}[1]{\label{eq:#1}}
\newcommand{\rf}[1]{(\ref{eq:#1})}

\newcommand{\hs}[1]{\hspace{#1 mm}}
\newcommand{\sm}[2]{\textstyle{\frac{#1}{#2}}\displaystyle}
\newcommand{\shalf}{\textstyle{\frac{1}{2}}\displaystyle}

\newcommand{\hg}{\widehat{g}}

\newcommand{\mub}{\bar{\mu}}

\newcommand{\mmb}{\mu \bar{\mu}}
\newcommand{\bz}{\bar{z}}
\newcommand{\bZ}{\bar{Z}}
\newcommand{\Zm}{Z_{\mu}}
\newcommand{\Zbm}{\bar{Z}_{\bar{\mu}}}
\newcommand{\lbm}{\lambda_{\mu}}
\newcommand{\lbbm}{\bar{\lambda}_{\bar{\mu}}}
\newcommand{\Sigm}{\Sigma_{\mu}}
\newcommand{\zbz}{(z, \bar{z})}
\newcommand{\ZBZ}{(Z, \bar{Z})}
\newcommand{\paz}{\partial_z}
\newcommand{\paZ}{\partial_Z}
\newcommand{\pabz}{\partial_{\bar{z}}}
\newcommand{\pabZ}{\partial_{\bar{Z}}}

\newcommand{\mes}[2]{\textstyle {\frac{d \bar{#1} 
\wedge d#2}{2i}}\displaystyle \ }

\newcommand{\bd}[1]{\, \raisebox{-3.8mm}{${\line(0,1){26}}_{ \ #1}$} }
\newcommand{\NPB}[1]{Nucl.\ Phys.\ {\bf B#1}}

\newcommand{\PLB}[1]{Phys.\ Lett.\ {\bf B#1}}

\newcommand{\CMP}[1]{Comm.\ Math.\ Phys.\ {\bf #1}}

\newtheorem{thm}{Theorem}[section]
\newtheorem{coro}[thm]{Corollary}

\newtheorem{prop}[thm]{Proposition}

\theoremstyle{definition}
\newtheorem{defi}[thm]{Definition}

\theoremstyle{remark}
\newtheorem{rem}[thm]{Remark}

\numberwithin{equation}{section}
\begin{document}

\title{The role of the Beltrami parametrization of complex
structures in 2-d Free Conformal Field Theory}

\author{Serge LAZZARINI} 

\address{Centre de Physique Th\'eorique-CNRS Luminy, Case 907,\\
F-13288 Marseille Cedex, France\\
and also Universit\'e de la M\'editerran\'ee.}

\email{sel@cpt.univ-mrs.fr}

\indent

\indent

\begin{abstract}
This talk gives a review on how complex geometry and a 
Lagrangian formulation of 2-d conformal field theory are deeply 
related. In particular, how the use of the
Beltrami parametrization of complex structures on a compact Riemann
surface fits perfectly with the celebrated locality principle of field theory,
the latter requiring the use infinite dimensional spaces.
It also allows a direct application of the local index theorem for families of
elliptic operators due to J.-M. Bismut, H. Gillet and C. Soul\'e.
The link between determinant line
bundles equipped with the Quillen's metric 
and the so-called holomorphic factorization property will be addressed
in the case of free spin $j$ b-c systems or more generally of 
free fields with values sections of
a holomorphic vector bundles over a compact Riemann surface.
\end{abstract}

\dedicatory{Talk given at the Colloque de G\'eom\'etrie Complexe,\\
 29 june - 3 July 1998, \\
Universit\'e de Paris 7, France\\[2mm]
published in\\[1mm]
G\'eom\'etrie Complexe II,\\ Aspects contemporains dans les
 math\'ematiques et la physique,\\
F.~Norguet \& S.~Ofman (Eds),\\
 Actualit\'es scientifiques et industrielles, Hermann, Paris, 2004}

\maketitle

\sect{Introduction}

In the last decade now, both string field and bidimensional conformal
field theories have enjoyed a large popularity in physics as well as in
mathematics, for a mathematical reading see e.g. 
\cite{Bo} and references therein. 
On the mathematical side, it seems rather remarkable how
some problems of these special classes of low dimensional field
theories helped to shed some new light on the classical geometry of
Riemann surfaces. Meanwhile, a refined version of the Atiyah-Singer
index theorem for 
families due to Bismut, Freed, Gillet and Soul\'e has been completed
\cite{BF,BGS,B,BGV}.

However, on the physical side, it is fair to note that quantization
schemes over moduli space have been widely preferred to the
conventional schemes (Feynman) grounded on some locality principle.
In particular, one of the main features of the bidimensional
conformal field theories is the so-called holomorphic-antiholomorphic
factorization property. 
Moreover, derivations of the factorization property from conventional
field theory scheme are so far limited to infinite dimensional spaces.

\indent

In this talk, we shall report on how the holomorphic factorization
property can be derived for free conformal fields on a compact
Riemann surface without boundary and of genus greater than one. It comes
from resummed renormalized perturbation theory for vacuum functionals
considered as defined on the infinite dimensional space of complex
structures on the Riemann surface (the well known Beltrami
parametrization) or on some holomorphic vector bundle
over it \cite{KLS,KLS',Sto90}. 
This provides a local version of the Belavin-Knizhnik theorem \cite{BK}
which is achieved by using the above mentioned local index theorem for
families by Bismut et al. This gives an important bridge with the locality principle of
the Euclidean version of field theory and the resummed renormalized
perturbation theory.

\indent

In Section 2, the free bosonic string will be recalled in the metric set
up. Section 3 describes the Beltrami parametrization of complex
structures and some of its advantages. Section 4 will concern the
application of the local index theorem in the version given By Bismut
et al. Finally some concluding remarks are gathered in section 5.

\sect{The classical String action}

Throughout this talk we will be concerned with the Euclidean framework.
In order to fix the idea one starts with the standard classical 
string action, \cite{DZ76,BdVH76,Pol81}, see also Bost \cite{Bo},
\begin{equation}
\Cl(X,g)\ =\ \sm{1}{8} \int_{\Sigma} d^2\!x \blp \sqrt{g}\,X\Delta(g)X\brp(x)
\ =\ - \sm{1}{8} \int_{\Sigma} d^2\!x \blp 
X \prt_{\alpha}\hg^{\alpha\beta}\prt_{\beta}X\brp(x)\ ,
\lbl{Sg}
\end{equation}
where $x=(x^1,x^2)$ denotes a set of local coordinates on the compact Riemann 
surface $\Sigma$ without boundary, $g$ is a Riemannian metric on this surface
with the associated Laplace-Beltrami operator, see e.g. \cite{Gol62},
acting on scalar functions, 
\begin{equation}
\Delta(g) = \frac{-1}{\sqrt{g}}\, \prt_{\alpha}\, \hg^{\alpha\beta}
\,\prt_{\beta}\ , \qquad \prt_{\alpha}= \frac{\prt}{\prt x^{\alpha}}\ ,
\lbl{Delta}
\end{equation}
acting on string coordinates $X(x)\in \mbox{\bf R}^D$. The second
integral in \rf{Sg} displays the Weyl invariant unimodular
non-degenerate quadratic form $\hg$ (density of metrics) 
related to the conformal class of the metric $g$ \cite{BBBC82,BBS86},
\begin{equation}
\hg_{\alpha\beta}\equiv
\frac{1}{\sqrt{g}}\,g_{\alpha\beta}\ ,\qquad 
\hg\equiv\det\hg = 1\ ,
\lbl{hg}
\end{equation}
showing the obvious invariance of \rf{Sg} under Weyl transformations.
See \cite{BBS86,BB87,Sto87,Bec88} for some historical background.

\sect{From metrics to complex structures}

In this section the relationship in two dimensions between conformal classes of
metrics and Beltrami parametrization of complex structure is
recalled.

\subsect{The smooth change of complex coordinates}

We shall right away introduce a complex analytic atlas 
with local coordinates $\zbz$ 
\footnote{We shall omit the index denoting the open set where these
coordinates are locally defined since all formulae will glue through 
holomorphic changes of coordinates.}
corresponding to the reference complex structure. We then parametrize
locally the metric $g$ according to \cite{Leh87},
\begin{equation}
ds^2 \ =\ g_{\alpha \beta} \, dx^{\alpha}dx^{\beta} \ =\ \rho_{z\bz} \,
|dz + {\mu_{\bz}}^z d\bz|^2 \ ,
\lbl{dsz}
\end{equation}
where ${\mu_{\bz}}^z\equiv\mu$ 
is the local representative of the Beltrami differential 
$(|\mu|<1)$ seen as a $(-1,1)$ conformal field, which parametrizes
the conformal class of the metric $g$ and $\rho_{z\bz}\equiv\rho>0$ 
is the coefficient of a positive real valued type (1,1) conformal field. 

Our compact Riemann surface $\Sigma$ without boundary being now 
endowed with the analytic atlas with local coordinates $\zbz$, 
let $(Z,\bZ)$ be the local coordinates of another holomorphic atlas
corresponding to the Beltrami differential $\mu$,
\begin{equation}
dZ \ =\ \lambda (dz + \mu d\bz) \qquad \Lra
\pabZ \ =\ \frac{\prtb - \mu \prt}{\bar{\lambda}(1 - \mmb)}
\lbl{dZ}
\end{equation}
where $\lambda\ =\ \prt_z Z\ \stackrel{\mbox{\tiny Def}}{\equiv}\,\prt Z$
is an integrating factor fulfilling
\footnote{From now on, we shall reserve $\prt$'s for 
$\prt\equiv\paz,\ \prtb\equiv\pabz$.}
\begin{equation}
(d^2\ =\ 0) \Llra 
(\prtb - \mu \prt) \ln \lambda \ = \ \prt \mu \ .
\lbl{lam}
\end{equation}
Solving the above Pfaff system \rf{dZ} is equivalent to solving locally 
the so-called Beltrami equation
\begin{equation}
(\prtb - \mu\prt)\, Z\ =\ 0\ .
\lbl{beltra}
\end{equation}
According to Bers, see e.g. \cite{Leh87}, the Beltrami equation
\rf{beltra} always admits as a solution a quasiconformal mapping
with dilatation coefficient $\mu$. One thus remarks that $Z$
is a (non-local) holomorphic functional of $\mu$ as well as the
integrating factor $\lambda$. However, the solution of the Beltrami equation 
define a smooth change of local complex coordinates $ \zbz \lra \ZBZ$ which 
preserves the orientation (the latter condition secures the 
requirement $|\mu|<1$), so that 
$\ZBZ$ defines a new system of 
complex coordinates with $Z\lra z$ when $|\mu|\lra 0$.

\subsect{The classical String action revisited}

In terms of the $\ZBZ$ complex coordinates which by virtue of \rf{dZ}
turns out to be isothermal coordinates for the metric $g$ by defining
the non-local metric
\begin{equation}
\rho_{Z\bZ}\equiv \frac{\rho}{\lambda\,\bar{\lambda}}\ .
\lbl{rZ}
\end{equation}
In particular, the quadratic form \rf{dsz}, the volume form, 
the scalar curvature and the scalar Laplacian, respectively write,
\begin{eqnarray}
ds^2 \ =\ g_{\alpha \beta} \, dx^{\alpha}dx^{\beta}\ =\ \rho_{Z\bZ}\,|dZ|^2,
&& \sqrt{g}\ =\ \rho\,(1-\mmb),\nn\\
&&{}\lbl{dsZ}\\[-2mm]
d^2\!x \sqrt{g} = \mes{Z}{Z}\rho_{Z\bZ},
&& \Delta(g)= \frac{-4}{\rho_{Z\bZ}}\,\paZ\pabZ,\nn
\end{eqnarray}
in such way that the classical string action \rf{Sg} now reads
\begin{equation}
S(X,\mu,\mub)\ =\ -\shalf \int_{\Sigma} \mes{Z}{Z} \lp X \paZ \pabZ X\rp\ZBZ\ ,
\lbl{SZ}
\end{equation}
which is nothing but the usual string action in the conformal gauge
\cite{Pol81} where the dependence in $\mu$ is hidden within the complex
coordinate $Z$ in an  {\em a priori} non-local way.
The latter is explicitly restored by going back to the $\zbz$ complex
coordinates \cite{BB87,Bec88} thanks to both \rf{dZ} and \rf{lam},
\begin{equation}
S(X,\mu,\mub)= -\shalf \int_{\Sigma}  \mes{z}{z} \shalf \lp
X(\prtb - \prt\mu)\frac{1}{1-\mmb}\,(\prt-\mub\prtb) X \rp\zbz,
\lbl{Sz}
\end{equation}
which turns out to be local in the complex structure $\mu$ thanks to
the disappearance of the non-local integrating factor upon using
eq.\rf{lam}. $\mu$ and $\mub$ are sources for the two components 
$\Theta$ (respectively $\bar{\Theta}$)  
\begin{equation}
\Theta\zbz\equiv \left. \frac{\delta\Cl}{\delta\mu\zbz}\right|_{\mu=\mub=0}
\ =\ -\shalf (\prt X)^2 \zbz\ ,
\lbl{emcl}
\end{equation}
of the energy-momentum tensor.

Roughly speaking the quantization \`a la Feynman of the string model 
yields the formal computation of the logarithm of the determinant of the
scalar Laplacian by a formal power series in $\mu,\mub,$,
$\Gamma(\mu,\mub)$, with
coefficients distributions, the correlation functions of the above
two components of the energy-momentum tensor. The question of
computing  
$$\Gamma(\mu,\mub)\ =\ \mbox{``$\ln\det \paZ\pabZ\ $ ''}$$
is addressed in the next section.

\sect{The application of the local index theorem for families}

\subsect{The geometrical data}

The {\em full} geometrical construction goes as follows. We first
choose a {\em prescribed} compact Riemann surface $(\Sigma,\zbz)$ without 
boundary where $\Sigma$ is a smooth compact surface without boundary and 
$(z,\bz)$ denotes a {\em fixed} local set of complex analytic coordinates 
on $\Sigma$. Now we are in a position to define
Beltrami differentials over $\Sigma$. That is, if $\kappa$ denotes 
the canonical holomorphic line bundle over $\Sigma$, we
consider the bundle of smooth $(-1,1)$-differentials over $\Sigma$
\begin{equation}
\kappa^{-1} \otimes \bar{\kappa} \lra \Sigma,
\lbl{1.1}
\end{equation}
and sections of this bundle are given by
\begin{equation}
\hat{\mu} \ = \ {\mu_{\bz}}^z \, d\bz \otimes \paz
\lbl{3}
\end{equation}
where ${\mu_{\bz}}^z \equiv \mu$ is the coefficient of the $(-1,1)$-differential
patching under holomorphic change of charts $(U,z) \rightarrow (V,w),\
w=w(z)$ as
\ceq
{\mu_{\bar{w}}}^w \ =\ \frac{w'}{\bar{w}'} \, {\mu_{\bz}}^z .
\cen
In order to obtain the space of Beltrami differentials parametrizing complex
structures on the surface $\Sigma$ we must impose some restrictive conditions
on the bundle \lbl{1.2}, i.e. on the coefficient 
$\mu$. We have $|{\mu_{\bz}}^z|<1$
and this absolute value is independent of coordinates, consequently the
$L_{\infty}$-norm of $\mu$ is well defined \cite{EE}.

Let $B$ be the infinite dimensional space of Beltrami differentials. 
This space has nice topological properties \cite{EE}~: it is a contractible 
complex manifold, convex and circled but non compact. Then any bundle over
$B$ is topologically trivial.

\subsect{The holomorphic family of compact Riemann surfaces}

Let $\pi: B \times \Sigma \lra B$ be the proper holomorphic map
of compact Riemann surfaces. For every $\mu \in B$, let
$\Sigm = \pi^{-1}\{\mu\}$ be the fiber over $\mu$, that is a Riemann
surface made by $\Sigma$ endowed with the complex structure given by $\mu$.
The local coordinates of the corresponding atlas will be denoted
$(\Zm,\Zbm)$ with
\begin{equation}
d\Zm \ = \ \lbm (dz + \mu d\bz) \ ,
\lbl{1.3}
\end{equation}
where $\lbm = \prt \Zm$ is an integrating factor which fulfills
\begin{equation}
(\prtb - \mu \prt) \ln \lbm \ = \ \prt \mu \ .
\lbl{1.5}
\end{equation}

\begin{rem} \normalsize
Note that the fiber $\Sigm$ is equivalent to the pair $(\Sigma,\mu)$.
More precisely, given a Beltrami differential $\mu$ on $\Sigma$ then 
surface $\Sigma$ is endowed with the new complex structure corresponding to
that $\mu$. So the new Riemann surface obtained is the fiber $\Sigm$. The
{\em reference} Riemann surface $(\Sigma,\zbz)$ is the 0-fiber
$\Sigma_0=(\Sigma,0)$.
\end{rem}

\indent

The local coordinates on $B \times \Sigma$ as smooth ($C^{\infty}$) trivial
bundle are $(\mu,\mub,z,\bz)$ if all fibers are seen to be modeled on 
$\Sigma$. The total differential is then
\begin{equation}
D \ = \ d + \prt_B + \prtb_B
\lbl{1.6}
\end{equation}
with $d = dz \prt_z + d\bz \prt_{\bz}$ and $\prt_B,\ \prtb_B$ are the
differentials
\footnote{The reader must keep in mind that $B$ is an infinite dimensional
manifold which will be treated formally the time being.}
according to the complex structure of $B$. We want to exhibit the
complex analytic structure on $B \times \Sigma$ induced by both those 
of $B$ and $\Sigma$ compatible with the $C^{\infty}$-structure. Indeed in the 
$C^{\infty}$-bundle structure it is defined by the local coordinates 
$(\mu,\mub,Z,\bZ)$ where the $Z$ coordinates are the generic coordinates on 
the fiber $\Sigma$. In other words, we obtain the local complex analytic 
coordinates on the fiber $\Sigm$ above $\mu$ by solving (locally)
the Beltrami equation (see eq.\rf{1.3})
\begin{equation}
(\prtb - \mu\prt ) Z \ = \ 0 \ ,
\lbl{1.7}
\end{equation}
and picking out solutions $\Zm$ holomorphic in $\mu$. 

Let $T(B \times \Sigma)$ be the tangent bundle over $B \times \Sigma$. 
We restrict ourselves to the vertical holomorphic tangent 
bundle $\kappa^{-1}_{\Sigma}$ which is a line bundle over $B \times \Sigma$
and its restriction to a fiber is isomorphic to the holomorphic tangent
bundle of this fiber. Notice that $\kappa_{\Sigma}$ is the canonical
holomorphic line bundle of {\em vertical} $(1,0)$-forms on $B \times
\Sigma$. Later on we shall be concerned with the {\em compact vertical cohomology
and the integration along the fiber} $\Sigma$ \cite{BT}. 

Now for every $\mu \in B$, due to the dimension of the fiber 
we can introduce a K\"ahler metric~\cite{KLS}
\begin{equation}
\rho_{\Zm\Zbm} \ = \ \frac{\rho_{z\bz}}{\lbm \lbbm} \ ,
\lbl{1.8'}
\end{equation}
on $\Sigm$ depending smoothly on $\mu$ and where $\rho_{z\bz} \equiv \rho$
is the K\"ahler metric on the 0-fiber.

Then the bundle $\kappa^{-1}_{\Sigma}$ is
endowed with the Hermitian metric
\begin{equation}
\rho_{Z\bZ} \ = \ \frac{\rho_{z\bz}}{\lambda \bar{\lambda}}.
\lbl{1.8}
\end{equation}

Moreover the induced norm on the Hermitian bundle 
$\kappa_{\Sigm}^{\otimes j} \otimes {{\bar{\kappa}_{\Sigm}}}^{\otimes
\bar{\jmath}}$
of $(j,\bar{\jmath})$-differentials $a(d\Zm)^j (d\Zbm)^{\bar{\jmath}}$ 
on $\Sigm$ is
\begin{eqnarray}
\| a \|^2 \ \stackrel{\rm Def}{\equiv} \ \| \alpha \|^2 
&=& \int_{\Sigma} 
\textstyle {\frac{d \Zbm \wedge \Zm}{2i}}\displaystyle \ | a |^2 \,
{\rho_{\Zm\Zbm}}^{1-j-\bar{\jmath}} 
\nonumber \\
&{}& \lbl{1.9} \\
&=& \int_{\Sigma} \mes{z}{z} (1 - \mmb) \, |\alpha|^2 \, 
\rho^{1-j-\bar{\jmath}} 
\nonumber
\end{eqnarray}
where thanks to the change of coordinates $\zbz \mapsto (\Zm,\Zbm)$
on $\Sigma$ 
\begin{equation}
a \ = \ \frac{\alpha}{\lbm^{j} \, \lbbm^{\bar{\jmath}}} \ .
\lbl{1.10}
\end{equation}
This last relation \rf{1.10} allows to consider the Hermitian bundle of
$(j,\bar{\jmath})$-differentials {\em over the $0$-fiber}.

\subsect{The determinant line bundle}

Our present task is to compute the first Chern class of the 
$\prtb$-determinant line bundle over $B$ by using the local form of the
Atiyah-Singer index theorem for families of elliptic operators given by
\cite{Bo,BF,BGS,B}. 

We are interested in the following {\em holomorphic} family of 
$\prtb$-operators over the compact Riemann surfaces $\Sigma$ fibers of
$\pi$~:
let $E= \kappa_{\Sigma}^{\otimes j}$ be the bundle whose restriction 
to the fiber $\Sigm$ is the bundle of smooth $(j,0)$-differentials on
this fiber and if $\Gamma(\Sigm,E)$ denotes the
smooth sections of the bundle $E$ over $B \times \Sigma$, we consider
\begin{equation}
\prtb_{\mu}:\ \Gamma(\Sigm,E) \lra
\Gamma(\Sigm,E \otimes \bar{\kappa}_{\Sigma}),
\lbl{fam}
\end{equation}
the operator $\prtb_{\mu}$ mapping $(j,0)$-differentials to 
$(j,1)$-differentials on $\Sigm$.

In particular if $c (d\Zm)^j$ is such a $(j,0)$-differential then by
the construction \rf{1.9} together with \rf{1.10} we get
\begin{eqnarray}
\|\prtb_{\mu} c\|^2 &=& \int_{\Sigma} 
\textstyle {\frac{d \Zbm \wedge \Zm}{2i}}\displaystyle \ 
|\prt_{\Zbm} c|^2 \, {\rho_{\Zm\Zbm}}^{-j}
\nonumber \\
&{}& \lbl{1.11} \\
&=& \int_{\Sigma} \mes{z}{z} \frac{1}{(1 - \mmb)} \,
|\mbox{\Large (} \prtb - \mu \prt - j(\prt \mu) \mbox{\Large )} \, \gamma|^2 
\, \rho^{-j} \ , \nonumber 
\end{eqnarray}
after use of
\begin{equation}
\prt_{\Zbm} \ = \ \frac{\prtb - \mu \prt}{\lbbm(1 - \mmb)}\ ,
\lbl{1.12}
\end{equation}
together with eq.\rf{1.5}, and where we have set by a local rescaling,
\begin{equation}
c\ =\ \frac{\gamma}{\lbm^j}\ ,
\end{equation}
in order to define the corresponding $(j,0)$-differential with respect
to the local complex coordinates $\zbz$.

\indent

\begin{prop}[\cite{KLS}] According to the formula \rf{1.11} the family of
(positive) Laplacians $\Delta_j$ on the compact Riemann surfaces $\Sigma$ 
is defined through
\begin{equation}
\| \prtb_{\mu} c \|^2 \ = \ <\!\gamma| \Delta_{j,\mu} \, \gamma\!> \ ,
\lbl{L}
\end{equation}
where $\Delta_{j,\mu} = \mathcal{T}_{j,\mu}^* \mathcal{T}_{j,\mu}$
with,
\begin{eqnarray}
\mathcal{T}_{j,\mu} &=& \frac{1}{1-\mmb} 
\mbox{\Large (} \prtb-\mu\prt-j(\prt\mu) \mbox{\Large )} \nonumber \\
&&{} \lbl{T}\\[1mm]
\mathcal{T}_{j,\mu}^* &=& \frac{-1}{\rho^{1-j}(1-\mmb)} \
\overline{\mbox{\Large(}\prtb-\mu\prt-(1-j)(\prt\mu) \mbox{\Large )}} 
\ \frac{1}{\rho^j}
\ = \ - \rho^{j-1} \, \bar{\mathcal{T}}_{1-j,\mu} \, \rho^{-j} \nonumber
\end{eqnarray}
where the adjoint is with respect to the norm of $(j,1)$-differentials, cf
eq.\rf{1.11}.
\end{prop}
Note that for $j=0$, the scalar Laplacian of the string action \rf{Sz}
is recovered.

\indent

\begin{prop}[\cite{KLS}]
Through the smooth change of local complex coordinates $(Z,\bZ) \mapsto \zbz$
the holomorphic family $\prtb_{\mu}$ \rf{fam} of $\prtb$-operators on compact 
Riemann surfaces $\Sigma$ is equivalent to the (only) smooth family 
$\mathcal{T}_j$ of $\prtb$-operators on the $0$-fiber $(\Sigma,\zbz)$,
\begin{equation}
\mathcal{T}_{j,\mu} :\ \Gamma(\Sigma,\kappa^{\otimes j}) \lra
\Gamma(\Sigma,\kappa^{\otimes j} \otimes \bar{\kappa}) .
\lbl{fam'}
\end{equation}
\end{prop}

\indent

\begin{rem} \normalsize
Therefore all computations can be performed on the reference Riemann surface
(i.e. the $0$-fiber).
\end{rem}

\indent

Let $\mathcal{L}$ denote the {\em determinant line bundle} of the family 
$\mathcal{T}_j$ over $B$ \cite{Q},
\begin{equation}
\mathcal{L} \ =\ \lambda({\rm Ker}\mathcal{T}_j)^* \otimes 
\lambda({\rm Ker}\mathcal{T}_{1-j})^*
\lbl{det}
\end{equation}
where we used ${\rm Ker}\mathcal{T}_j^* \simeq ({\rm Ker}\mathcal{T}_{1-j})^*$
thanks to the second formula in \rf{T}.

 It has a canonical {\em holomorphic} structure, see below,
even if the family $\mathcal{T}_j$  is smooth., cf eq.\rf{T}.
If $\Sigma$ denotes the fixed Riemann surface (0-fiber) then we have the 
following diagram 
\ceq
\begin{array}{ccccc}
\kappa^{-1}_{\Sigma} &                      & \mathcal{L} \\
\downarrow           &                      & \downarrow \\
B \times \Sigma      & \stackrel{\pi}{\lra} & B &
\stackrel{\stackrel{\mu}{\leftarrow}}{\lra} & \Sigma
\end{array}
\cen 

\indent

\begin{rem} \normalsize
Since the base $B$ is contractible the first Chern class of the
determinant line bundle
$\mathcal{L}$ is cohomologically trivial that is $H^2(B,{\bf R})=0$.
\end{rem}

\subsect{The Quillen metric on $\mathcal{L}$}

We shall choose a basis $\{\gamma_{m,\mu}\}$ (resp. 
$\{\beta_{p,\mu}\}$), $m = 1, \dots ,N_j$ (resp. $p~= ~1, \dots
,N_{1-j})$ in the kernel of $\mathcal{T}_{j,\mu}$ (resp. $\mathcal{T}_{1-j,\mu}$) 
{\em holomorphic} in $\mu$ 
\footnote{This can be done locally in $\mu$.}
and {\em independent} of $\rho$, with, according to the Riemann-Roch theorem
\ceq
N_j - N_{1-j} \ = \ (g-1)(2j-1) \ ,
\cen
if $\Sigma$ is of genus $g$.

By construction let $s$ be the non-vanishing holomorphic section
\begin{equation}
s\ =\ (\gamma_1 \wedge \ldots \wedge \gamma_{N_j})^{-1} \otimes
(\beta_1 \wedge \ldots \wedge \beta_{N_{1-j}})^{-1},
\lbl{sec}
\end{equation}
of $\mathcal{L}$, see eq.\rf{det}. Then

\indent

\begin{defi}[\cite{Q} or Bost in \cite{Bo}]
The Quillen metric on $\mathcal{L}$ is defined to be
\begin{equation}
\| s \|^2_Q \ =\ ({\rm det'}_{\zeta} \, \Delta_j) \, \| s \|^2_{L_2}
\lbl{Q}
\end{equation}
where the norm
\ceq
\| s \|^2_{L_2} = \frac{1}{\det<\gamma_m| \gamma_n>_{1\leq m,n\leq N_j}}
\,\frac{1}{\det<\beta_p|\beta_q>_{1\leq p,q\leq N_{1-j}}},
\cen
all metrics $<\ |\ >$ (on the $0$-fiber) are associated with
$ds^2 = \rho |dz + \mu d\bz|^2$, and ${\rm det'}_{\zeta}$ stands for the 
$\zeta$-regularized determinant
\ceq
\ln {\rm det'}_{\zeta} \, \Delta_j \ = \ - \frac{d}{ds} \bd{s=0} \,
\frac{1}{\Gamma(s)} \int_0^{\infty} t^{s-1}dt \left({\rm Tr} \,e^{-t\Delta_j} 
- N_j \right) \ .
\cen
\end{defi}

\subsect{The local index theorem for the family $\mathcal{T}_j$}

In our case of interest the hypotheses are the following~:
Let $\pi: b\times\Sigma\lra B$ be a smooth proper holomorphic map of Riemann surfaces.
Assume that $\pi$ is locally K\"ahler, i.e. there is an open covering
$\mathcal{U}$ of $B$ such that if $U \in \mathcal{U},\ \pi^{-1}(U)$ admits a 
K\"ahler metric (whose restriction to the fiber $\Sigm,\ \mu \in U$ may
differ from the K\"ahler metric \rf{1.8'}). 
Let $E$ be as above the Hermitian bundle equipped with the
Hermitian metric induced by \rf{1.8}. Let $\mathcal{K},\ \mathcal{E}$ be the
curvatures of the holomorphic Hermitian metric on ${\kappa^{-1}_{\Sigma}}$
and $E$ respectively. Then the theorem states that

\indent

\begin{thm}[\cite{Bo,BGS}] Under the above hypotheses, the curvature of the
connection associated with the Quillen metric on the determinant line bundle 
$\mathcal{L}$ is the component of degree 2 in the following form on $B$
\begin{equation}
2i\pi \biggl[
\pi_* {\rm Td}(-\mathcal{K}/2i\pi) \, {\rm Tr} [\exp(-\mathcal{E}/2i\pi)] 
\biggr]^{(2)} ,
\lbl{thm1}
\end{equation}
where $\pi_* : \Omega^*_{vc}(B \times \Sigma) \lra \Omega^{*-2}(B)$ is
the ``integration along the fiber" in the compact vertical cohomology which
commutes with the exterior differentiation $d_B$ on the base \cite{BT}.
\end{thm}

\indent

Consider now for a the quantity
\begin{equation}
\Gamma(\rho,\mu,\mub) - \Gamma(\rho,0,0), \qquad \Gamma(\rho,\mu,\mub)
= \shalf \ln \| s \|^2_Q,
\lbl{gam}
\end{equation}
for a given section \rf{sec} locally holomorphic in $\mu$.
Since all bundles are line bundles we can express the equality of differential 
forms on $B$ in the following 

\indent

\begin{coro}[e.g. \cite{Bo}] The $1$-st Chern class of $\mathcal{L}$ locally 
represented by
\begin{equation}
c_1(\mathcal{L}) \ =\ - \sm{1}{2i\pi} \prtb_B \prt_B \ln \| s \|^2_Q = -
\sm{1}{2i\pi} \prtb_B \prt_B \lp \Gamma(\rho,\mu,\mub) - \Gamma(\rho,0,0)\rp,
\lbl{curv}
\end{equation}
equals
\begin{equation}
c_1(\mathcal{L}) \ =\ - [ \pi_* \, {\rm Td}(\kappa^{-1}_{\Sigma}) 
{\rm Ch}(E) ]^{(2)} \ =\ - \sm{C_j}{12} \, 
\pi_* c_1(\kappa^{-1}_{\Sigma})^2 \ ,
\lbl{thm2} 
\end{equation}
where the constant $C_j = 6j(j-1)+1$.
\end{coro}

\indent

For the sake of definiteness, let us consider an analytic submanifold $B_t
\subset B$ parametrized by a finite dimensional holomorphic family $\mu_t$,
with $\mu_0=0$, where $t$ denotes a set of local coordinates of a finite 
dimensional analytic manifold $S$, (e.g. a neighbourhood of the origin
in ${\bf C}^n$). Following the construction made in \cite {W} let us 
introduce the map
\begin{equation}
\begin{array}{rcl}
\Phi :\ S \times \Sigma & \lra & B \times \Sigma \\
(t,\zbz) & \longmapsto & (\mu_t,Z_{\mu_t}\zbz)
\end{array}
\lbl{1.19}
\end{equation}
which is holomorphic in $t$ and a $C^{\infty}$-diffeomorphism of $\zbz$, so 
that the coordinate $Z_{\mu_t} \equiv Z_t$ built up from the Beltrami equation
\rf{1.7} is chosen to be holomorphic in $t$.

Thus, to the complex analytic structure defined by the natural coordinates
$(t,\bar{t},Z,\bZ)$ there corresponds the following splitting of the total
differential \rf{1.6}
\begin{equation}
D\ =\ \mathcal{D} + \bar{\mathcal{D}} \ =\ d + d_t + d_{\bar{t}}
\lbl{1.20}
\end{equation}
with
\begin{equation}
\mathcal{D} \ =\ DZ \paZ + Dt \, \prt_{{t}|Z}, \ \ 
\mathcal{D}^2 = \mathcal{D} \bar{\mathcal{D}} + \bar{\mathcal{D}} \mathcal{D} =
\bar{\mathcal{D}}^2 = 0,
\lbl{1.21}
\end{equation}
where $\prt_{{t}|Z}$ denotes partial derivatives with respect to
the $t$ coordinates at constant $Z,\bZ$.

Now the first Chern class $c_1(\kappa^{-1}_{\Sigma}) \in H^2(B_t
\times \Sigma,{\bf R})$ of the holomorphic vertical line bundle over $B_t
\times \Sigma$ which corresponds to the Hermitian metric \rf{1.8}
is locally represented by the $D$-closed form
\footnote{In the following, the product of differentials is to be understood
as the usual wedge product.}
on $B_t \times \Sigma$ given by
\begin{eqnarray}
(-2i\pi) c_1(\kappa^{-1}_{\Sigma}) & = & \mathcal{K} 
\ =\ \bar{\mathcal{D}} \mathcal{D} \ln\!\rho_{Z\bZ} \nonumber \\[3mm]
&=& (D\bZ \, \pabZ + D\bar{t} \, \prt_{\bar{t}|Z})
(DZ \, \paZ + Dt \, \prt_{t|Z}) \ln\!\rho_{Z\bZ} \lbl{1.30}
\\[3mm]
&=& \left( D\bZ DZ \paZ \pabZ + 
D\bar{t} DZ \, \prt_{\bar{t}|Z} \paZ
+ D\bZ Dt \, \pabZ \prt_{t|Z}  \right.\nn\\[3mm]
&&\left.\qquad +\ D\bar{t} Dt \, \prt_{\bar{t}|Z} 
\prt_{t|Z} \right) \ln\!\rho_{Z\bZ} \ .\nn
\end{eqnarray}
Performing the smooth change of local complex coordinates 
$(t,\bar{t},Z,\bZ) \mapsto (t,\bar{t},z,\bz)$
through the expression \rf{1.3} of $dZ$ here seen as a section over $B_t$,
we get
\begin{equation}
\begin{array}{ll}
(-2i\pi) c_1(\kappa^{-1}_{\Sigma}) \ = & \left[ (d\bZ + d\bar{t} \,
\prt_{t} \bZ)(dZ + dt \, \prt_{t} Z) \paZ \pabZ 
+ d\bar{t} \, (dZ + dt \, \prt_{t}Z) \prt_{\bar{t}|Z} \paZ \right. \\[3mm]
& + \left. (d\bZ + d\bar{t} \, \prt_{\bar{t}} \bZ) dt \, \prt_{t|Z} 
\pabZ + d\bar{t} \, dt \, \prt_{\bar{t}|Z} \prt_{t|Z} \right] 
\ln\!\rho_{Z\bZ} 
\end{array}
\lbl{1.31}
\end{equation}
Using now the chain rule
\ceq
\prt_{t} \ = \ \prt_{t|Z} + (\prt_{t} Z) \paZ
\cen
where $\prt_{t} = \prt_{t|z}$ eq.\rf{1.31} becomes
\begin{equation}
(-2i\pi) c_1(\kappa^{-1}_{\Sigma}) \ = \ \left[ d\bZ dZ \paZ \pabZ + d\bZ dt \,
\prt_{t} \pabZ + d\bar{t} \, dZ \, \prt_{\bar{t}} \paZ 
\right] \ln\!\rho_{Z\bZ} 
\lbl{1.32}
\end{equation}
The ``square" of this first Chern class belongs directly to the vertical
compact cohomology class $H_{vc}^4(B_t \times \Sigma,{\bf R})$
\begin{equation}
c_1(\kappa^{-1}_{\Sigma})^2 \ =\ \sm{1}{2\pi^2} \
d\bZ dZ d\bar{t} \, dt \: | \prt_{t} \pabZ \ln\!\rho_{Z\bZ} |^2  .
\lbl{1.33}
\end{equation}

\indent

Now the curvature \rf{curv} gives rise to an obstruction to the
holomorphic factorization property, namely
\ceq
\prtb_B \prt_B \lp \Gamma(\rho,\mu,\mub) - \Gamma(\rho,0,0)\rp \neq 0.
\cen
Happily, one has

\begin{thm}[\cite{KLS}] The square of the first Chern class of the vertical
holomorphic line bundle over $B_t \times \Sigma$ is 
\begin{equation}
c_1(\kappa^{-1}_{\Sigma})^2 \ =\ \sm{i}{\pi^2} 
d_{\bar{t}} \, d_{t} \, \chi(\rho,\mu,\mub),
\lbl{1.35}
\end{equation}
{\em where $\chi(\rho,\mu,\mub)$ is a section in
$\Gamma(B_t,\Omega^{1,1}(\Sigma,{\bf R}))$ and depends locally on 
$\rho,\ \mu,\ \mub$}.
\end{thm}
\begin{proof} See \cite{KLS} formulas (37,38).
The expression of $\chi$ is given in \cite{KLS} formula (22) (but actually
comes from elsewhere \cite{KLT})
\begin{eqnarray}
\chi(\rho,\mu,\mub) &=& \frac{d \bar{z} 
\wedge dz}{2i}
\mbox{\Huge (}
\mu \, (\mathcal{R} - R) \ + \ c.c. \
- \ \frac{1}{1-\mmb} \mbox{\LARGE [} (\prt + \Gamma) \mu (\prtb + \bar{\Gamma}) 
\mub \nonumber \\ 
&&{} \\
\lbl{1.36}
&& \hs{20} - \shalf \mub
\mbox{\Large (} (\prt + \Gamma) \mu \mbox{\Large )}^2 - \shalf \mu
\mbox{\Large (} (\prtb + \bar{\Gamma}) \mub \mbox{\Large )}^2 
\mbox{{\LARGE ]}{\Huge )}} \ , \nonumber
\end{eqnarray}
where $\mathcal{R}$ is the coefficient of a {\em holomorphic projective connection}
\cite{G} and $\Gamma = \prt \ln\!\rho,\ \bar{\Gamma} = \prtb \ln\!\rho, 
\ R = \prt \Gamma - \shalf \Gamma^2$.
\end{proof}
Note that $\chi(\rho,0,0) = 0$.

\indent

\begin{coro} One reaches the conclusion that 
\begin{equation}
\prtb_B \prt_B \lp \Gamma(\rho,\mu,\mub) - \Gamma(\rho,0,0) +
\sm{C_j}{12\pi} \pi_* \chi(\rho,\mu,\mub)\rp = 0,
\end{equation}
and there is no longer any dependence on the Weyl factor $\rho$,
\begin{equation}
\prt_{\rho}\lp\Gamma(\rho,\mu,\mub) - \Gamma(\rho,0,0) +
\sm{C_j}{12\pi} \pi_* \chi(\rho,\mu,\mub)\rp = 0.
\end{equation}
\end{coro}
Thus holomorphic factorization follows but it is far from
being unique. It depends on the existence of the section \rf{sec} of
the determinant line bundle, covariant under diffeomorphisms a problem
which has not been addressed.

\indent

\begin{coro} The first Chern class of the determinant line bundle 
$\mathcal{L}$ is then locally expressed as a $\prtb_B \prt_B$-exact form
\begin{equation}
c_1(\mathcal{L}) \ =\ \sm{C_j}{12i\pi^2} \
d_{\bar{t}} \, d_{t} \, \pi_* \chi(\rho,\mu,\mub),
\lbl{1.40}
\end{equation}
which can be viewed as another representative of this first Chern class
different from that given by the Quillen metric.
\end{coro}

\indent

This construction can be extended to free fields considered as sections of a
holomorphic bundle \cite{KLS'} where besides the Beltrami differential
parametrizing the complex structures on the Riemann surface, there is
the $(0,1)$-component of a connection parametrizing the complex
structures of the holomorphic bundle. One reaches the same conclusion
as above.
 
\sect{Concluding remarks} 

The holomorphic factorization property illustrates the deep relationship
between the locality principle of Euclidean field theory and the refined
version of the index theorem for elliptic families. The introduction
of metrics which cannot be avoided at present, positivity being the
main ingredient allowing to sum up the perturbative series, remains
compatible with the analyticity.

Accordingly to previous works \cite{KLS,KLS'}, 
the lecturer together with M.~Knecht and
R.~Stora resulted in a striking geometrical situation from an application of
the local index theorem for families of $\prtb$-operators parametrized by
Beltrami differentials. The outcome was that the first Chern class of the
determinant line bundle over the space of Beltrami differentials, locally
expressed with the aid of Quillen's metric given by
\begin{equation}
\prtb_B \prt_B \ln \| s \|_Q^2 \ =\ - \sm{C_j}{6\pi} \
\prtb_B \prt_B \int_{\Sigma}\chi,
\lbl{0}
\end{equation}
with $\chi \in \Omega^{1,1}(\Sigma,{\bf R})$ a $(1,1)$-form on the typical fiber
$\Sigma$ and $s$ a non-vanishing section of this determinant bundle. 
Surprisingly this $(1,1)$-form depends locally on the Beltrami 
differentials. 

The resulting problem amounts to finding a geometrical reason 
why the identity \rf{1.35} or \rf{1.40} holds. What is first remarkable is that 
even the 1-st Chern class $c_1(\mathcal{L})$ must indeed be $d_B$-exact (thanks 
to the topological triviality) but it is more~: it is $\prtb_B
\prt_B$-exact of a local quantity~! 
So one ought to think of a kind of {\em local holomorphic} triviality. 
But the computation carried out in \cite{KLS} shows that the identity 
\rf{1.40} has a global meaning. 
According to the definition \rf{gam} of what can be considered as
a resummation of the perturbative series, one has to face a kind of
interpolating formula between two Quillen metrics, and the expression
for $\pi_* \chi$ resembles a Bott-Chern transgression formula.

\indent

\noindent
{\bf Acknowledgements.} The author is grateful to the organizers of this
Colloquium for the opportunity to illustrate with this example the
deep interplay between geometry and field theory.

\bibliographystyle{amsalpha}

\begin{thebibliography}{A}

\bibitem{Bo} J.-B.~Bost ; Bourbaki Seminar, Ast\'erique Vol. {\bf
        152-153} (1986).\\
        D.S.~Freed ; ``On Determinant Line Bundles", in ``Mathematical
        Aspects of String Theory", S.T. Yau ed., World Scientific (1987).\\
        L.A. Takhtajan ; ``Uniformization, Local Index Theorem and
        Geometry of the Moduli Spaces of Riemann Surfaces and Vector Bundles",
        Proceedings of Symposium in Pure Mathematics, {\bf 49} (1989) 581-596.

\bibitem{BF} J.-M.~Bismut, D.S.~Freed ; \CMP{107} (1986) 103.

\bibitem{BGS} J.-M. Bismut, H.~Gillet, and C.~Soul\'e.
\newblock Analytic torsion and holomorphic determinant bundles, {I}, {II} {\&}
  {III}.
\newblock {\em \CMP{115}}, (1988), pages 49, 79 {\&} 301.

\bibitem{B} J.-M.~Bismut ; Inv. Math. {\bf 83} (1986) 91 ; 
        Math. Ann. {\bf287} (1990) 495 ; Inv. Math. {\bf 99} (1990) 59 ;
        ``Superconnexions, indice local des familles, d\'eterminant de
        la cohomologie et m\'etriques de Quillen", Orsay preprint
        91-04 ; ``Holomorphic Families of Immersions and Higher
        Analytic Torsion Forms", Ast\'erique Vol. {\bf 244} 1997.

\bibitem{BGV} N. Berline, E. Getzler and M. Vergne, ``Heat Kernels and
        Dirac Operators", Grundlehren des Mathematischen Wissenschasft {\bf
        298}, Springer-Verlag, Berlin Heidelberg 1992.

\bibitem{KLS} M.~Knecht, S.~Lazzarini, R.~Stora ; ``On Holomorphic
        Factorization for Free Conformal Fields", \PLB{262} (1991) 25.

\bibitem{KLS'} M.~Knecht, S.~Lazzarini, R.~Stora ; ``On Holomorphic
        Factorization for Free Conformal Fields~II", \PLB{273} (1991) 63.

\bibitem{Sto90} R.~Stora.
\newblock On {L}agrangian two dimensional conformal models.
\newblock {\em Progr. Theor. Phys. Suppl. {\bf 102}}, pages 373--386, 1990.
\newblock Lectures given at the Yukawa International Seminar on ``Common Trends
  in Mathematics and Quantum Field Theories'', 10-19 May 1990, Kyoto, Japan.

\bibitem{BK} A.A. Belavin and V.G. Knizhnik.
        \newblock Algebraic geometry and the geometry of quantum strings.
        \newblock {\em \PLB{168}}, pages 201--206, (1986).\\
        A.A. Belavin and V.G. Knizhnik.
        \newblock Complex geometry and the theory of quantum strings.
        \newblock {\em Sov. Phys. JETP {\bf 64}}, pages 214--228, (1986).

\bibitem{DZ76} S.~Deser and B.~Zumino.
\newblock A complete action for the spinning string.
\newblock {\em \PLB{65}}, pages 369--373, (1976).

\bibitem{BdVH76} L.~Brink, P.~Di Vecchia, and P.~Howe.
\newblock A locally supersymmetric and reparametrization invariant action for
  the spinning string.
\newblock {\em \PLB{65}}, pages 471--474, (1976).

\bibitem{Pol81} A.M. Polyakov.
\newblock {Q}uantum geometry of bosonic strings.
\newblock {\em \PLB{103}}, pages 207--210, (1981).

\bibitem{Gol62} S.L. Goldberg.
\newblock {\em Curvature and homology}.
\newblock Dover Publications, New York, 1962.
\newblock Republication 1982.

\bibitem{BBBC82}
G.~Bandelloni, C.~Becchi, A.~Blasi, and R.~Collina.
\newblock Local approach to dilatation invariance.
\newblock {\em \NPB{197}}, pages 347--364, (1982).

\bibitem{BBS86}
L.~Baulieu, C.~Becchi, and R.~Stora.
\newblock ``{O}n the covariant quantization of the free bosonic string''.
\newblock {\em \PLB{180}}, pages 55--60, (1986).

\bibitem{BB87}
L.~Baulieu and M.~Bellon.
\newblock ``{B}eltrami parametrization in string theory''.
\newblock {\em Phys. Lett. {\bf B196}}, page 142, (1987).

\bibitem{Sto87}
R.~Stora.
\newblock Alg{\`e}bres diff{\'e}rentielles en th{\'e}orie des champs.
\newblock {\em Ann. Inst. Fourier {\bf 37}, 4}, pages 235--245, (1987).

\bibitem{Bec88}
C.M. Becchi.
\newblock ``{O}n the covariant quantization of the free string: the conformal
  structure''.
\newblock {\em Nucl. Phys. {\bf B304}}, page 513, (1988).

\bibitem{Leh87}
O.~Lehto.
\newblock {\em Univalent functions and {T}eichm{\"u}ller spaces}.
\newblock Graduate Texts in Mathematics. Springer-Verlag, Berlin, 1987.

\bibitem{EE} C.J.~Earle, J.~Eells ; J. Diff. Geom. {\bf 3} (1969) 19.
\bibitem{BT} R.~Bott, L.W.~Tu ; ``Differential Forms in Algebraic
Topology", Springer-Verlag (1982).

\bibitem{AT97}
E.~Aldrovandi and L.A. Takhtajan.
\newblock Generating functional in {CFT} and effective action for
  two-dimensional quantum gravity on higher genus {R}ie\-mann sur\-fa\-ces.
\newblock {\em Commun. Math. Phys. {\bf 188}}, pages 29--67, (1997).
\newblock {\tt hep-th/9606163}.
\bibitem{Q} D.~Quillen ; Funct. Anal. Appl. {\bf 19} (1985) 31. \\
        D.~Quillen ; Topology {\bf 24} (1985) 89.
\bibitem{W} S.A.~Wolpert ; Invent. Math. {\bf 85} (1986) 119.
\bibitem{KLT} M.~Knecht, S.~Lazzarini, F.~Thuillier ; \PLB{251} (1990) 279.

\bibitem{G} R.~Gunning~; ``Lectures on Riemann Surfaces'', Princeton
Univ. Press~(1966). 
\end{thebibliography}

\end{document}